\documentclass[%
reprint,
superscriptaddress,
showpacs,
 amsmath,amssymb,
 aps,
 pra,
 floats,
]{revtex4-1}

\usepackage[latin1]{inputenc}
\usepackage{graphicx}
\usepackage{dcolumn}
\usepackage{bm}
\usepackage{units}
\usepackage{color}
\usepackage{float}

\begin{document}

\preprint{APS/123-QED}

\title{Experimental stress--strain analysis of tapered silica optical fibers with nanofiber waist}%

\author{S. Holleis}
\author{T. Hoinkes}
\author{C. Wuttke}
\author{P. Schneeweiss}
\author{A. Rauschenbeutel}
\affiliation{%
 Vienna Center for Quantum Science and Technology,\\
 TU Wien -- Atominstitut, Stadionallee 2, 1020 Vienna, Austria
}%

\date{\today}

\begin{abstract}
We experimentally determine tensile force--elongation diagrams of tapered optical fibers with a nanofiber waist. The tapered optical fibers are produced from standard silica optical fibers using a heat and pull process. Both, the force--elongation data and scanning electron microscope images of the rupture points indicate a brittle material. Despite the small waist radii of only a few hundred nanometers, our experimental data can be fully explained by a nonlinear stress--strain model that relies on material properties of macroscopic silica optical fibers. This is an important asset when it comes to designing miniaturized optical elements as one can rely on the well-founded material characteristics of standard optical fibers. Based on this understanding, we demonstrate a simple and non-destructive technique that allows us to determine the waist radius of the tapered optical fiber. We find excellent agreement with independent scanning electron microscope measurements of the waist radius.

\end{abstract}
\maketitle



Glass fibers are among the most versatile inventions of the last century with applications reaching from fiber-reinforced materials as used in construction~\cite{Bunsell05} to optical data transmission in global telecommunication networks~\cite{Kaminow08}. In recent years, so-called tapered optical fibers (TOFs) have  received growing attention~\cite{Brambilla10}, both as optical components, e.g., for coupling light into micro- and nano-optical components, for sensing, as well as for the controlled coupling of light and matter at or near the TOF surface~\cite{Vetsch10,Morrissey13,Swaim13,Liebermeister14}. Beyond their optical properties, which have been extensively studied in the past, it is important to understand and control the mechanical properties of these devices for many of these applications.

Here, we study the mechanical response of silica TOFs with a nanofiber waist when exposed to tensile stress. Qualitatively, we observe a brittle stress--strain behavior and no constriction of the two fiber ends when the fiber has been ruptured apart. Previous studies on the mechanical properties of such ultra-thin silica structures are divided over the importance of size effects in the nanoscopic domain, e.g., suggesting a decrease or an increase of Young's modulus compared to bulk values~\cite{Dikin03,Silva05,Ni05,brambilla}. In our study, even for the smallest investigated waist radius of only $160$~nm, the recorded stress--strain diagrams match those of macroscopic optical fibers~\cite{mallinder,krause,glaesemann}. As a consequence, in the parameter range considered here, it is justified to assume the well-studied mechanical properties of standard optical fibers when designing nano-optical elements. Based on this understanding, we provide a practical, non-destructive in-situ method to determine the TOF waist radius. The results obtained in this way are in very good agreement with independent radius measurements using a scanning electron microscope.

\begin{figure}
	\includegraphics{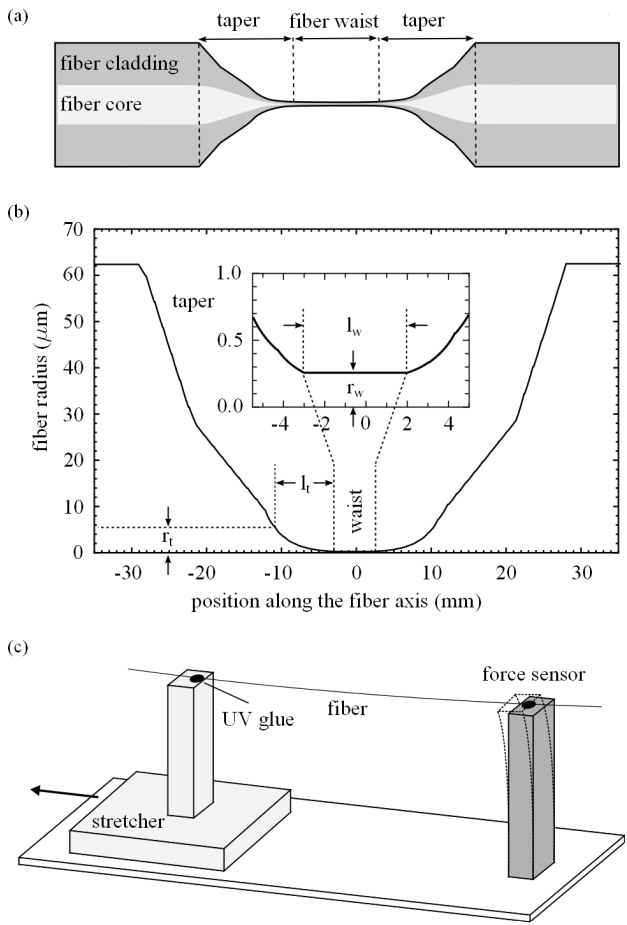}
	\caption{(a) Schematic of a tapered optical fiber (TOF) showing the different regions: unprocessed fiber, taper region, and nanofiber waist. (b) Example radius profile of a TOF introducing the relevant quantities $l_{\rm t}$, $r_{\rm t}$, $l_{\rm w}$, and $r_{\rm w}$. The inset shows a zoom of the thinnest section of the TOF. (c) Sketch of the experimental setup: One end of the TOF is attached to a precision translation stage (stretcher) while the other end is fixed to a force sensor. The stretcher can be moved to the left in a controlled manner. In this way, the fiber is elongated until it ruptures while the force sensor records the tensile force during the entire process.}
	\label{fig:principle} 
\end{figure}

The TOFs we study in our experiments are produced from commercial optical fibers (SM800, Fibercore) using a heat and pull process~\cite{Birks92,warken}. A schematic of such a cylindrically symmetric TOF is shown in Fig.~\ref{fig:principle}(a). The TOF consists of three different regions: the unprocessed fiber with a radius of 62.5~$\mu$m, the taper, and the fiber waist or nanofiber section. The radius varies over more than two orders of magnitude for the different regions of the TOF. This is illustrated in more detail in Fig.~\ref{fig:principle}(b), where we show an example radius profile of a TOF. For the stress--strain experiments presented here, only the thinnest parts of the TOF are relevant as sections with larger radius will experience negligible strain. These relevant sections are the cylindrical nanofiber waist with nominal length $l_{\rm w}$ and nominal radius $r_{\rm w}$, and the connecting onset of the taper on each side of the waist. The radius of the TOF in such a taper onset varies exponentially, i.e., $r(x)\propto \exp (-\chi x)$, and reaches a nominal value $r_{\rm t}$ within a nominal length $l_{\rm t}$. 

We study TOFs with nominal waist radii $r_{\rm w}$ between $160$~nm and $500$~nm, and nominal waist lengths $l_{\rm w}$ of $2$, $5$, and $10$~mm. For all fibers discussed here, the radius increases with the exponent $\chi=0.389\,/$mm. The nominal quantities $r_{\rm t}$ and $l_{\rm t}$ of the taper onset are chosen to be compatible with technical constrains of the fabrication process. Due to imperfections of the fabrication, the actual shape of the TOF deviates from its nominal shape. In particular, the actual waist radius, $r_{\rm w}^{\rm act}$, fluctuates by about $\pm 10$~\% around $r_{\rm w}$~\cite{stiebeiner}. As a trait of the fabrication process, the relative deviations from the other nominal dimensions are much smaller~\cite{warken}.

\begin{figure}
	\includegraphics{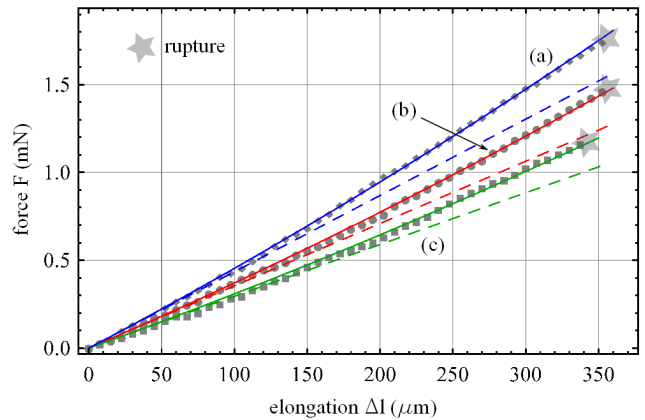}
	\caption{(Color online) Gray symbols: measured tensile force as a function of elongation for three fibers, (a), (b), and (c). The colored solid lines on top are fits to the respective data based on a nonlinear stress--strain model, see main text. The only free fit parameters are the waist radii which have been found to be $299$ nm (a), $268$ nm (b) and $242$ nm (c), respectively. In order to visualize the deviation from linearity, we also show the prediction of the model for these radii when setting the nonlinear term to zero, see dashed lines. }
	\label{fig:fitsem} 
\end{figure}

For our experiments, we use a force sensor (KD78 with strain gauge amplifier GSV-2TDS-DI, Measuring Equipment) which is able to measure forces in the millinewton range. A sketch of the experimental setup is given in Fig.~\ref{fig:principle}(c). The two ends of the TOF are fixed to the force sensor and to a precision translation stage (stretcher) with small drops of UV-curing glue. The position of the stretcher is computer controlled such that we can automatically move it away from the force sensor at a constant velocity of 25 $\mu$m/s. During this process, the fiber is elongated until it ruptures. The force sensor is read out continuously with a sampling rate of 100 Hz and records a force-over-time trace. By multiplying the time axis with the constant velocity of the stretcher, we obtain a force-over-elongation trace as shown in Fig.~\ref{fig:fitsem} for three example TOFs.

Here, the measured tensile force $F(\Delta{l})$ is displayed as gray symbols for three different TOFs, (a), (b), and (c). We typically stretch the TOFs for a few 100 $\mu$m before they rupture. The values of the maximum tensile force lie in the 1--2~mN range and depend on the waist radius of the respective nanofiber. The measured curves show typical brittle behavior~\cite{Beer04} and are essentially linear with a slight superlinear contribution.

In order to quantitatively describe the experimental data, we follow the approach presented by E.~Suhir~\cite{suhir} and calculate the total elongation of the fiber using
\begin{eqnarray}
\Delta{l}=\int\limits_{-l/2}^{l/2}\epsilon(x)dx~,
\label{eq:elongation}
\end{eqnarray}
where $l=l_{\rm w}+2l_{\rm t}$, and $\epsilon(x)$ describes the local strain along the fiber. Furthermore, we employ the following nonlinear stress--strain relationship:
\begin{eqnarray}
\sigma=E\epsilon(1+\frac{1}{2}\alpha\epsilon)~,
\label{eq:stresstrain}
\end{eqnarray}
where the first term describes the linear contribution (Hooke's law) and the second term accounts for a quadratic contribution. The quantities in Eq.~(\ref{eq:stresstrain}) are the tensile stress $\sigma$, Young's modulus $E$, and the weight $\alpha$ of the quadratic term. In the following, we assume the average values for standard silica optical fibers: $E=72$ GPa and $\alpha=6$ as obtained experimentally in~\cite{mallinder,krause,glaesemann}. The stress $\sigma(x)$ at any position $x$ of the tapered fiber is then calculated according to 
\begin{eqnarray}
\sigma(x)=\frac{F}{\pi r^2(x)}=\sigma_{\rm w} \frac{r_{\rm w}^2}{r^2(x)}~,
\label{eq:stressforce}
\end{eqnarray}
where $F$ is the measured tensile force, $\pi r(x)^2$ is the cross-sectional area of the fiber at the position $x$, $r_{\rm w}$ is the waist radius, and $\sigma_{\rm w}=\frac{F}{\pi r_{\rm w}^2}$ is the stress experienced by the waist. From Eqs.~(\ref{eq:stresstrain}) and (\ref{eq:stressforce}) one obtains~\cite{suhir}
\begin{eqnarray}
\epsilon(x)=\frac{1}{\alpha}\left[\left(1+2\alpha \frac{F}{\pi E} \frac{1}{r(x)^2}\right)^{\frac{1}{2}}-1\right],
\label{eq:strain}
\end{eqnarray} 
which we insert into Eq.~(\ref{eq:elongation}). This yields the total elongation as a function of the applied tensile force, $\Delta{l(F)}$. In order to model the measured force--elongation data, we only take into account the thinnest parts of the TOF, i.e., the cylindrical nanofiber waist section and the onsets of the two taper transitions with the dimensions as introduced before. Thus, the radius profile $r(x)$ in Eq.~(\ref{eq:strain}) takes the form
		\begin{equation}
		r(x)=\left\{
		\begin{matrix}
			r_{\rm w} e^{-\chi (x + l_{\rm w}/2)}  & -(l_{\rm t} + l_{\rm w}/2) \leq x < -l_{\rm w}/2 \\
			r_{\rm w} & -l_{\rm w}/2 \leq x < + l_{\rm w}/2  \\
			r_{\rm w} e^{\chi (x - l_{\rm w}/2)}  & l_{\rm w}/2 \leq x < l_{\rm w}/2 + l_{\rm t}~. \\
		\end{matrix}
		\right.
		\end{equation}
For the three example fibers shown in Fig.~\ref{fig:fitsem}, we use the nominal values, $l_{\rm t}=7.794$~mm, $r_{\rm t}= 5.407$~$\mu$m, and $l_{\rm w}=1.983$~mm, respectively, in our model. Note that for $r_{\rm t}\gg r_{\rm w}$, the exact value of $r_{\rm t}$ is not important due to the negligible contribution of the thick part of the taper onset to the integral in Eq.~(\ref{eq:elongation}). Since the actual waist radius, $r_{\rm w}^{\rm act}$, exhibits the largest deviations from its nominal value, we chose it as the only free fit parameter. A least-squares minimization yields the fit curves shown as solid lines in Fig.~\ref{fig:fitsem}, matching very well the measurements. The obtained waist radii are $r_{\rm w}^{\rm act}=299$ nm (a), $268$ nm (b), and $242$ nm (c), respectively. About 60~\% of the total elongation of the TOF is contributed by the two taper onsets that enclose the nanofiber waist. In order to visualize the importance of the quadratic term in the stress--stain relationship in Eq.~(\ref{eq:stresstrain}), we show the prediction of the model for the determined actual radii when setting the nonlinear term to zero [$\alpha=0$ in Eq.~(\ref{eq:stresstrain})], see dashed lines. The linear model performs well for low strain. However, for higher strain, close to the rupture point of nanofibers, the nonlinear model is required in order to obtain good agreement with the experimental results.

\begin{figure*}
	\includegraphics{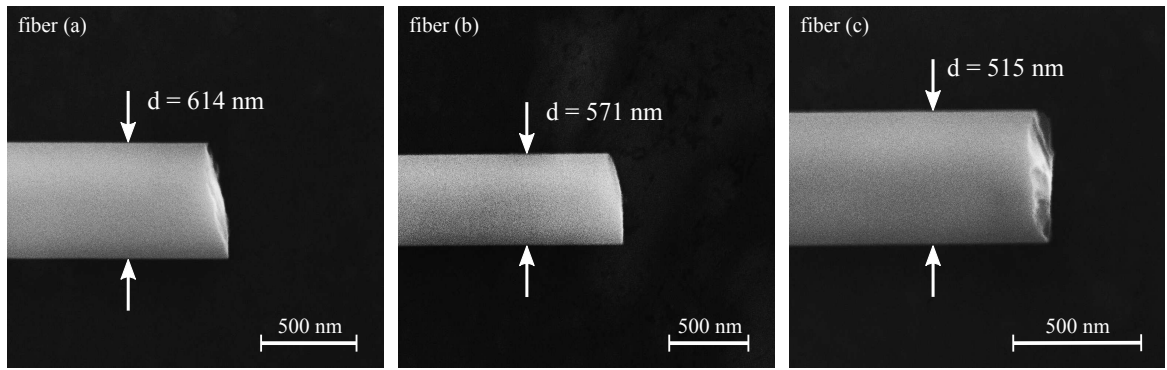}
	\caption{Scanning electron microscope images of the points of rupture of the three tapered optical fibers from Fig.~\ref{fig:fitsem}. Rough rupture surfaces are clearly visible. The radii measured with the SEM agree within 5\% with the fitted radii obtained from the force--elongation measurements. SEM parameters: electron acceleration voltage: 20 kV, working distance: 10.5 mm.}
	\label{fig:semfibers} 
\end{figure*}

In order to independently check the determined values for $r_{\rm w}^{\rm act}$, we examine the three fibers from Fig.~\ref{fig:fitsem} with a scanning electron microscope (SEM, FEI Quanta 200 FEGSEM). The corresponding SEM images are shown in Fig.~\ref{fig:semfibers}. The values of $r_{\rm w}^{\rm act}$ deviate by less than 5~\% from the respective SEM results. In addition, the SEM images show the points of rupture of the nanofibers. In contrast to previous work by Brambilla and Payne~\cite{brambilla}, in which corresponding SEM images are interpreted to show clean cuts, we observe rough rupture surfaces. In agreement with Ref.~\cite{brambilla}, we observe no constriction in the close vicinity of the point of rupture. This is expected for a brittle material~\cite{Campbell12} and, thus, in agreement with the measured force--elongation data in Fig.~\ref{fig:fitsem}. 

We now measure force--elongation curves for sets of TOFs with different nominal waist radii and lengths which are specified in the first two columns of Tab.~\ref{tab:strains}. From each measurement, we determine $r_{\rm w}^{\rm act}$ as above and note the rupture force $F^{\rm rupt}$. This allows us to infer the strain and stress at the waist when it ruptures, ${\epsilon}_{\rm w}^{\rm rupt}$ and ${\sigma}_{\rm w}^{\rm rupt}$, respectively. For each combination of $r_{\rm w}$ and $l_{\rm w}$, five TOFs are analyzed and the mean value as well as the standard deviation of ${\epsilon}_{\rm w}^{\rm rupt}$ and ${\sigma}_{\rm w}^{\rm rupt}$ are given in Tab.~\ref{tab:strains}. The maximum strain and stress scatter by a few percent within each set of TOFs and are remarkably constant across all sets. Averaging over all samples, we obtain ${\epsilon}_{\rm w}^{\rm rupt}=(7.3\pm0.5$)\% and ${\sigma}_{\rm w}^{\rm rupt}=(6.4\pm0.5$)~GPa. This agrees well with literature values for standard silica optical fibers, see last line of Tab.~\ref{tab:strains}. 

Based on these results, we suggest a simple non-destructive method to determine the actual waist radius, $r_{\rm w}^{\rm act}$ of a TOF in situ. To these ends, the force--elongation relationship of the TOF in the regime well below the rupture force is measured and $r_{\rm w}^{\rm act}$ is extracted using the model discussed above. In order to test this method, several TOFs are strained up to an increase in length of 3\% and then relieved. This procedure is repeated ten times for each TOF (see~\cite{kobayashi,sudarshanam,bertholds} for similar measurements on standard optical fibers). We observe agreement of consecutive force traces for increasing and for decreasing load, i.e., no hysteresis is apparent. Moreover, all ten pairs of force traces overlap within the precision of our measurements, indicating the absence of fatigue of the fiber material for our conditions and confirming the non-destructive character of the method. The waist radii obtained with our approach agree within 3\% with the ones obtained from SEM measurements. The combination of simplicity and relatively high accuracy of our method offers an interesting alternative to existing schemes for the radius measurement of nanofibers~\cite{Warken04,Sumetsky06b,Wiedemann10}.
\begin{table}
\begin{tabular}{|c|c||c|c|}
			\hline
			 $r_{\rm w}$ (nm) & $l_{\rm w}$ (mm) & $\epsilon_{\rm w}^{\rm rupt}$ (\%) & $\sigma_{\rm w}^{\rm rupt}$ (GPa)
		 \\
			\hline \hline
			160 & 2 & $7.5\pm0.3$ & $6.6\pm0.3$ \\ \hline
			250 & 2 & $7.6\pm0.3$ & $6.8\pm0.3$ \\ \hline
			250 & 5 & $6.9\pm0.2$ & $6.0\pm0.3$ \\ \hline
			250 & 10 & $6.6\pm0.1$ & $5.7\pm0.1$ \\ \hline
			350 & 2 & $7.4\pm0.2$ & $6.5\pm0.2$ \\ \hline
			500 & 2 & $7.9\pm0.2$ & $7.0\pm0.2$ \\ \hline \hline
			\multicolumn{2}{|c||}{all samples} & $7.3\pm0.5$ & $6.4\pm0.5$ \\ \hline \hline
			\multicolumn{2}{|c||}{standard silica fiber} & {6.5~\cite{mallinder,krause}} & {5.6~\cite{kurkjian}} \\ \hline
		\end{tabular}		\caption{Rupture strain and stress in the TOF waist: $r_{\rm w}$ and $l_{\rm w}$ are the nominal waist radii and waist lengths of the nanofibers. For each combination of $r_{\rm w}$ and $l_{\rm w}$, a set of five TOFs has been examined. The average over all samples is in reasonable agreement with literature values for standard silica optical fibers, see last line.}
		\label{tab:strains}
\end{table}

In summary, we measured force--elongation diagrams of tapered optical fibers with a nanofiber waist. Our data indicate a brittle material and are well explained using a nonlinear stress--strain model that relies on the material properties of standard silica optical fibers. The waist radii obtained from the fit of the force--elongation measurements are in excellent agreement with independent measurements using a scanning electron microscope. Images of ruptured nanofibers showed a rough rupture surface and no sign of constriction. In repeated force--elongation measurements on the same TOF, we observe no hysteresis, thereby enabling a non-destructive determination of the TOF waist radius. 

An absence of a length-scale dependence of material properties down to a system size on the order of optical wavelengths would be of large technical relevance. For all data presented in this work, including these for TOFs with only $160$~nm waist radius, the measured force--elongation relationships are well explained assuming the macroscopic values of Young's modulus and of the nonlinearity parameter $\alpha$. Recent experiments have observed that also other macroscopic material parameters remain valid for systems with wavelength-scale dimensions~\cite{Wuttke13b}. This holds promising perspectives for, e.g.,~the precise the design of the mechanical and optical properties of miniaturized photonic components as well as for fundamental studies with optical nanofibers. 
 

Financial support by the Austrian Science Fund (FWF, SFB NextLite Project No. F 4908-N23 and SFB FoQus Project No. F 4017-N23) is gratefully acknowledged. The scanning electron microscope imaging has been carried out using facilities at the University Service Centre for Transmission Electron Microscopy, Vienna University of Technology, Austria.

\bibliography{StressStrain}

\begin{thebibliography}{28}%
\makeatletter
\providecommand \@ifxundefined [1]{%
 \@ifx{#1\undefined}
}%
\providecommand \@ifnum [1]{%
 \ifnum #1\expandafter \@firstoftwo
 \else \expandafter \@secondoftwo
 \fi
}%
\providecommand \@ifx [1]{%
 \ifx #1\expandafter \@firstoftwo
 \else \expandafter \@secondoftwo
 \fi
}%
\providecommand \natexlab [1]{#1}%
\providecommand \enquote  [1]{``#1''}%
\providecommand \bibnamefont  [1]{#1}%
\providecommand \bibfnamefont [1]{#1}%
\providecommand \citenamefont [1]{#1}%
\providecommand \href@noop [0]{\@secondoftwo}%
\providecommand \href [0]{\begingroup \@sanitize@url \@href}%
\providecommand \@href[1]{\@@startlink{#1}\@@href}%
\providecommand \@@href[1]{\endgroup#1\@@endlink}%
\providecommand \@sanitize@url [0]{\catcode `\\12\catcode `\$12\catcode
  `\&12\catcode `\#12\catcode `\^12\catcode `\_12\catcode `\%12\relax}%
\providecommand \@@startlink[1]{}%
\providecommand \@@endlink[0]{}%
\providecommand \url  [0]{\begingroup\@sanitize@url \@url }%
\providecommand \@url [1]{\endgroup\@href {#1}{\urlprefix }}%
\providecommand \urlprefix  [0]{URL }%
\providecommand \Eprint [0]{\href }%
\providecommand \doibase [0]{http://dx.doi.org/}%
\providecommand \selectlanguage [0]{\@gobble}%
\providecommand \bibinfo  [0]{\@secondoftwo}%
\providecommand \bibfield  [0]{\@secondoftwo}%
\providecommand \translation [1]{[#1]}%
\providecommand \BibitemOpen [0]{}%
\providecommand \bibitemStop [0]{}%
\providecommand \bibitemNoStop [0]{.\EOS\space}%
\providecommand \EOS [0]{\spacefactor3000\relax}%
\providecommand \BibitemShut  [1]{\csname bibitem#1\endcsname}%
\let\auto@bib@innerbib\@empty
\bibitem [{\citenamefont {Bunsell}\ and\ \citenamefont
  {Renard}(2005)}]{Bunsell05}%
  \BibitemOpen
  \bibfield  {author} {\bibinfo {author} {\bibfnamefont {A.~R.}\ \bibnamefont
  {Bunsell}}\ and\ \bibinfo {author} {\bibfnamefont {J.}~\bibnamefont
  {Renard}},\ }\href@noop {} {\emph {\bibinfo {title} {Fundamentals of Fibre
  Reinforced Composite Materials}}}\ (\bibinfo  {publisher} {Institute of
  Physics Publishing},\ \bibinfo {year} {2005})\BibitemShut {NoStop}%
\bibitem [{\citenamefont {I.~P.~Kaminow}\ and\ \citenamefont
  {Willner}(2008)}]{Kaminow08}%
  \BibitemOpen
  \bibinfo {editor} {\bibfnamefont {T.~L.}\ \bibnamefont {I.~P.~Kaminow}}\ and\
  \bibinfo {editor} {\bibfnamefont {A.~E.}\ \bibnamefont {Willner}},\ eds.,\
  \href@noop {} {\emph {\bibinfo {title} {Optical Fiber Telecommunications}}}\
  (\bibinfo  {publisher} {Elsevier},\ \bibinfo {year} {2008})\BibitemShut
  {NoStop}%
\bibitem [{\citenamefont {Brambilla}(2010)}]{Brambilla10}%
  \BibitemOpen
  \bibfield  {author} {\bibinfo {author} {\bibfnamefont {G.}~\bibnamefont
  {Brambilla}},\ }\href {http://stacks.iop.org/2040-8986/12/i=4/a=043001}
  {\bibfield  {journal} {\bibinfo  {journal} {J. of Opt.}\ }\textbf {\bibinfo
  {volume} {12}},\ \bibinfo {pages} {043001} (\bibinfo {year}
  {2010})}\BibitemShut {NoStop}%
\bibitem [{\citenamefont {Vetsch}\ \emph {et~al.}(2010)\citenamefont {Vetsch},
  \citenamefont {Reitz}, \citenamefont {Sagu\'e}, \citenamefont {Schmidt},
  \citenamefont {Dawkins},\ and\ \citenamefont {Rauschenbeutel}}]{Vetsch10}%
  \BibitemOpen
  \bibfield  {author} {\bibinfo {author} {\bibfnamefont {E.}~\bibnamefont
  {Vetsch}}, \bibinfo {author} {\bibfnamefont {D.}~\bibnamefont {Reitz}},
  \bibinfo {author} {\bibfnamefont {G.}~\bibnamefont {Sagu\'e}}, \bibinfo
  {author} {\bibfnamefont {R.}~\bibnamefont {Schmidt}}, \bibinfo {author}
  {\bibfnamefont {S.~T.}\ \bibnamefont {Dawkins}}, \ and\ \bibinfo {author}
  {\bibfnamefont {A.}~\bibnamefont {Rauschenbeutel}},\ }\href {\doibase
  10.1103/PhysRevLett.104.203603} {\bibfield  {journal} {\bibinfo  {journal}
  {Phys. Rev. Lett.}\ }\textbf {\bibinfo {volume} {104}},\ \bibinfo {pages}
  {203603} (\bibinfo {year} {2010})}\BibitemShut {NoStop}%
\bibitem [{\citenamefont {Morrissey}\ \emph {et~al.}(2013)\citenamefont
  {Morrissey}, \citenamefont {Deasy}, \citenamefont {Frawley}, \citenamefont
  {Kumar}, \citenamefont {Prel}, \citenamefont {Russell}, \citenamefont
  {Truong},\ and\ \citenamefont {Nic~Chormaic}}]{Morrissey13}%
  \BibitemOpen
  \bibfield  {author} {\bibinfo {author} {\bibfnamefont {M.~J.}\ \bibnamefont
  {Morrissey}}, \bibinfo {author} {\bibfnamefont {K.}~\bibnamefont {Deasy}},
  \bibinfo {author} {\bibfnamefont {M.}~\bibnamefont {Frawley}}, \bibinfo
  {author} {\bibfnamefont {R.}~\bibnamefont {Kumar}}, \bibinfo {author}
  {\bibfnamefont {E.}~\bibnamefont {Prel}}, \bibinfo {author} {\bibfnamefont
  {L.}~\bibnamefont {Russell}}, \bibinfo {author} {\bibfnamefont {V.~G.}\
  \bibnamefont {Truong}}, \ and\ \bibinfo {author} {\bibfnamefont
  {S.}~\bibnamefont {Nic~Chormaic}},\ }\href {\doibase 10.3390/s130810449}
  {\bibfield  {journal} {\bibinfo  {journal} {Sensors}\ }\textbf {\bibinfo
  {volume} {13}},\ \bibinfo {pages} {10449} (\bibinfo {year}
  {2013})}\BibitemShut {NoStop}%
\bibitem [{\citenamefont {Swaim}\ \emph {et~al.}(2013)\citenamefont {Swaim},
  \citenamefont {Knittel},\ and\ \citenamefont {Bowen}}]{Swaim13}%
  \BibitemOpen
  \bibfield  {author} {\bibinfo {author} {\bibfnamefont {J.~D.}\ \bibnamefont
  {Swaim}}, \bibinfo {author} {\bibfnamefont {J.}~\bibnamefont {Knittel}}, \
  and\ \bibinfo {author} {\bibfnamefont {W.~P.}\ \bibnamefont {Bowen}},\ }\href
  {\doibase http://dx.doi.org/10.1063/1.4829659} {\bibfield  {journal}
  {\bibinfo  {journal} {Appl. Phys. Lett.}\ }\textbf {\bibinfo {volume}
  {103}},\ \bibinfo {eid} {203111} (\bibinfo {year} {2013})}\BibitemShut
  {NoStop}%
\bibitem [{\citenamefont {Liebermeister}\ \emph {et~al.}(2014)\citenamefont
  {Liebermeister}, \citenamefont {Petersen}, \citenamefont {Münchow},
  \citenamefont {Burchardt}, \citenamefont {Hermelbracht}, \citenamefont
  {Tashima}, \citenamefont {Schell}, \citenamefont {Benson}, \citenamefont
  {Meinhardt}, \citenamefont {Krueger}, \citenamefont {Stiebeiner},
  \citenamefont {Rauschenbeutel}, \citenamefont {Weinfurter},\ and\
  \citenamefont {Weber}}]{Liebermeister14}%
  \BibitemOpen
  \bibfield  {author} {\bibinfo {author} {\bibfnamefont {L.}~\bibnamefont
  {Liebermeister}}, \bibinfo {author} {\bibfnamefont {F.}~\bibnamefont
  {Petersen}}, \bibinfo {author} {\bibfnamefont {A.~v.}\ \bibnamefont
  {Münchow}}, \bibinfo {author} {\bibfnamefont {D.}~\bibnamefont {Burchardt}},
  \bibinfo {author} {\bibfnamefont {J.}~\bibnamefont {Hermelbracht}}, \bibinfo
  {author} {\bibfnamefont {T.}~\bibnamefont {Tashima}}, \bibinfo {author}
  {\bibfnamefont {A.~W.}\ \bibnamefont {Schell}}, \bibinfo {author}
  {\bibfnamefont {O.}~\bibnamefont {Benson}}, \bibinfo {author} {\bibfnamefont
  {T.}~\bibnamefont {Meinhardt}}, \bibinfo {author} {\bibfnamefont
  {A.}~\bibnamefont {Krueger}}, \bibinfo {author} {\bibfnamefont
  {A.}~\bibnamefont {Stiebeiner}}, \bibinfo {author} {\bibfnamefont
  {A.}~\bibnamefont {Rauschenbeutel}}, \bibinfo {author} {\bibfnamefont
  {H.}~\bibnamefont {Weinfurter}}, \ and\ \bibinfo {author} {\bibfnamefont
  {M.}~\bibnamefont {Weber}},\ }\href {\doibase
  http://dx.doi.org/10.1063/1.4862207} {\bibfield  {journal} {\bibinfo
  {journal} {Appl. Phys. Lett.}\ }\textbf {\bibinfo {volume} {104}},\ \bibinfo
  {eid} {031101} (\bibinfo {year} {2014})}\BibitemShut {NoStop}%
\bibitem [{\citenamefont {Dikin}\ \emph {et~al.}(2003)\citenamefont {Dikin},
  \citenamefont {Chen}, \citenamefont {Ding}, \citenamefont {Wagner},\ and\
  \citenamefont {Ruoff}}]{Dikin03}%
  \BibitemOpen
  \bibfield  {author} {\bibinfo {author} {\bibfnamefont {D.~A.}\ \bibnamefont
  {Dikin}}, \bibinfo {author} {\bibfnamefont {X.}~\bibnamefont {Chen}},
  \bibinfo {author} {\bibfnamefont {W.}~\bibnamefont {Ding}}, \bibinfo {author}
  {\bibfnamefont {G.}~\bibnamefont {Wagner}}, \ and\ \bibinfo {author}
  {\bibfnamefont {R.~S.}\ \bibnamefont {Ruoff}},\ }\href {\doibase
  http://dx.doi.org/10.1063/1.1527971} {\bibfield  {journal} {\bibinfo
  {journal} {J. of Appl. Phys.}\ }\textbf {\bibinfo {volume} {93}},\ \bibinfo
  {pages} {226} (\bibinfo {year} {2003})}\BibitemShut {NoStop}%
\bibitem [{\citenamefont {Silva}\ \emph {et~al.}(2006)\citenamefont {Silva},
  \citenamefont {Tong}, \citenamefont {Yip},\ and\ \citenamefont
  {Van~Vliet}}]{Silva05}%
  \BibitemOpen
  \bibfield  {author} {\bibinfo {author} {\bibfnamefont {E.~C. C.~M.}\
  \bibnamefont {Silva}}, \bibinfo {author} {\bibfnamefont {L.}~\bibnamefont
  {Tong}}, \bibinfo {author} {\bibfnamefont {S.}~\bibnamefont {Yip}}, \ and\
  \bibinfo {author} {\bibfnamefont {K.~J.}\ \bibnamefont {Van~Vliet}},\ }\href
  {\doibase 10.1002/smll.200500311} {\bibfield  {journal} {\bibinfo  {journal}
  {Small}\ }\textbf {\bibinfo {volume} {2}},\ \bibinfo {pages} {239} (\bibinfo
  {year} {2006})}\BibitemShut {NoStop}%
\bibitem [{\citenamefont {Ni}\ \emph {et~al.}(2006)\citenamefont {Ni},
  \citenamefont {Li},\ and\ \citenamefont {Gao}}]{Ni05}%
  \BibitemOpen
  \bibfield  {author} {\bibinfo {author} {\bibfnamefont {H.}~\bibnamefont
  {Ni}}, \bibinfo {author} {\bibfnamefont {X.}~\bibnamefont {Li}}, \ and\
  \bibinfo {author} {\bibfnamefont {H.}~\bibnamefont {Gao}},\ }\href {\doibase
  http://dx.doi.org/10.1063/1.2165275} {\bibfield  {journal} {\bibinfo
  {journal} {Appl. Phys. Lett.}\ }\textbf {\bibinfo {volume} {88}},\ \bibinfo
  {eid} {043108} (\bibinfo {year} {2006})}\BibitemShut {NoStop}%
\bibitem [{\citenamefont {Brambilla}\ and\ \citenamefont
  {Payne}(2009)}]{brambilla}%
  \BibitemOpen
  \bibfield  {author} {\bibinfo {author} {\bibfnamefont {G.}~\bibnamefont
  {Brambilla}}\ and\ \bibinfo {author} {\bibfnamefont {D.~N.}\ \bibnamefont
  {Payne}},\ }\href@noop {} {\bibfield  {journal} {\bibinfo  {journal} {Nano
  Lett.}\ }\textbf {\bibinfo {volume} {9}},\ \bibinfo {pages} {831} (\bibinfo
  {year} {2009})}\BibitemShut {NoStop}%
\bibitem [{\citenamefont {Mallinder}\ and\ \citenamefont
  {Proctor}(1964)}]{mallinder}%
  \BibitemOpen
  \bibfield  {author} {\bibinfo {author} {\bibfnamefont {F.~P.}\ \bibnamefont
  {Mallinder}}\ and\ \bibinfo {author} {\bibfnamefont {B.~A.}\ \bibnamefont
  {Proctor}},\ }\href@noop {} {\bibfield  {journal} {\bibinfo  {journal} {Phys.
  Chem. Glasses}\ }\textbf {\bibinfo {volume} {5, 4}} (\bibinfo {year}
  {1964})}\BibitemShut {NoStop}%
\bibitem [{\citenamefont {Krause}\ \emph {et~al.}(1979)\citenamefont {Krause},
  \citenamefont {Testardi},\ and\ \citenamefont {Thurston}}]{krause}%
  \BibitemOpen
  \bibfield  {author} {\bibinfo {author} {\bibfnamefont {J.~T.}\ \bibnamefont
  {Krause}}, \bibinfo {author} {\bibfnamefont {L.~R.}\ \bibnamefont
  {Testardi}}, \ and\ \bibinfo {author} {\bibfnamefont {R.~N.}\ \bibnamefont
  {Thurston}},\ }\href@noop {} {\bibfield  {journal} {\bibinfo  {journal}
  {Phys. Chem. Glasses}\ }\textbf {\bibinfo {volume} {20, 6}} (\bibinfo {year}
  {1979})}\BibitemShut {NoStop}%
\bibitem [{\citenamefont {Glaesemann}\ \emph {et~al.}(1988)\citenamefont
  {Glaesemann}, \citenamefont {Gulati},\ and\ \citenamefont
  {Helfinstine}}]{glaesemann}%
  \BibitemOpen
  \bibfield  {author} {\bibinfo {author} {\bibfnamefont {G.~S.}\ \bibnamefont
  {Glaesemann}}, \bibinfo {author} {\bibfnamefont {S.~T.}\ \bibnamefont
  {Gulati}}, \ and\ \bibinfo {author} {\bibfnamefont {J.~D.}\ \bibnamefont
  {Helfinstine}},\ }\href@noop {} {\bibfield  {journal} {\bibinfo  {journal}
  {Opt. Fiber Comm.}\ }\textbf {\bibinfo {volume} {1}} (\bibinfo {year}
  {1988})}\BibitemShut {NoStop}%
\bibitem [{\citenamefont {{Birks}}\ and\ \citenamefont {{Li}}(1992)}]{Birks92}%
  \BibitemOpen
  \bibfield  {author} {\bibinfo {author} {\bibfnamefont {T.~A.}\ \bibnamefont
  {{Birks}}}\ and\ \bibinfo {author} {\bibfnamefont {Y.~W.}\ \bibnamefont
  {{Li}}},\ }\href {\doibase 10.1109/50.134196} {\bibfield  {journal} {\bibinfo
   {journal} {J. Lightwave Technol.}\ }\textbf {\bibinfo {volume} {10}},\
  \bibinfo {pages} {432} (\bibinfo {year} {1992})}\BibitemShut {NoStop}%
\bibitem [{\citenamefont {Warken}(2007)}]{warken}%
  \BibitemOpen
  \bibfield  {author} {\bibinfo {author} {\bibfnamefont {F.}~\bibnamefont
  {Warken}},\ }\emph {\bibinfo {title} {Ultradünne Glasfasern als Werkzeug zur
  Kopplung von Licht und Materie}},\ \href@noop {} {Ph.D. thesis},\ \bibinfo
  {school} {Rheinische Friedrich-Wilhelms-Universität Bonn} (\bibinfo {year}
  {2007})\BibitemShut {NoStop}%
\bibitem [{\citenamefont {Stiebeiner}\ \emph {et~al.}(2010)\citenamefont
  {Stiebeiner}, \citenamefont {Garcia-Fernandez},\ and\ \citenamefont
  {Rauschenbeutel}}]{stiebeiner}%
  \BibitemOpen
  \bibfield  {author} {\bibinfo {author} {\bibfnamefont {A.}~\bibnamefont
  {Stiebeiner}}, \bibinfo {author} {\bibfnamefont {R.}~\bibnamefont
  {Garcia-Fernandez}}, \ and\ \bibinfo {author} {\bibfnamefont
  {A.}~\bibnamefont {Rauschenbeutel}},\ }\href@noop {} {\bibfield  {journal}
  {\bibinfo  {journal} {Opt. Express}\ }\textbf {\bibinfo {volume} {18, 22}},\
  \bibinfo {pages} {22677} (\bibinfo {year} {2010})}\BibitemShut {NoStop}%
\bibitem [{\citenamefont {Beer}\ \emph {et~al.}(2004)\citenamefont {Beer},
  \citenamefont {Johnston},\ and\ \citenamefont {DeWolf}}]{Beer04}%
  \BibitemOpen
  \bibfield  {author} {\bibinfo {author} {\bibfnamefont {F.~P.}\ \bibnamefont
  {Beer}}, \bibinfo {author} {\bibfnamefont {E.~R.}\ \bibnamefont {Johnston}},
  \ and\ \bibinfo {author} {\bibfnamefont {J.~T.}\ \bibnamefont {DeWolf}},\
  }\href@noop {} {\emph {\bibinfo {title} {Mechanics Of Materials}}}\ (\bibinfo
   {publisher} {McGraw-Hill},\ \bibinfo {year} {2004})\BibitemShut {NoStop}%
\bibitem [{\citenamefont {Suhir}(1993)}]{suhir}%
  \BibitemOpen
  \bibfield  {author} {\bibinfo {author} {\bibfnamefont {E.}~\bibnamefont
  {Suhir}},\ }\href@noop {} {\bibfield  {journal} {\bibinfo  {journal} {Applied
  Optics}\ }\textbf {\bibinfo {volume} {32, 18}} (\bibinfo {year}
  {1993})}\BibitemShut {NoStop}%
\bibitem [{\citenamefont {Campbell}(2012)}]{Campbell12}%
  \BibitemOpen
  \bibinfo {editor} {\bibfnamefont {F.~C.}\ \bibnamefont {Campbell}},\ ed.,\
  \href@noop {} {\emph {\bibinfo {title} {Fatigue and Fracture: Understanding
  the Basics}}}\ (\bibinfo  {publisher} {ASM International},\ \bibinfo {year}
  {2012})\BibitemShut {NoStop}%
\bibitem [{\citenamefont {Kobayashi}(1978)}]{kobayashi}%
  \BibitemOpen
  \bibfield  {author} {\bibinfo {author} {\bibfnamefont {H.}~\bibnamefont
  {Kobayashi}},\ }\href@noop {} {\bibfield  {journal} {\bibinfo  {journal} {J.
  Appl. Phys.}\ }\textbf {\bibinfo {volume} {49}},\ \bibinfo {pages} {4776}
  (\bibinfo {year} {1978})}\BibitemShut {NoStop}%
\bibitem [{\citenamefont {Sudarshanam}\ and\ \citenamefont
  {Srinivasan}(1990)}]{sudarshanam}%
  \BibitemOpen
  \bibfield  {author} {\bibinfo {author} {\bibfnamefont {V.~S.}\ \bibnamefont
  {Sudarshanam}}\ and\ \bibinfo {author} {\bibfnamefont {K.}~\bibnamefont
  {Srinivasan}},\ }\href@noop {} {\bibfield  {journal} {\bibinfo  {journal}
  {Opt. Lett.}\ }\textbf {\bibinfo {volume} {15, 23}} (\bibinfo {year}
  {1990})}\BibitemShut {NoStop}%
\bibitem [{\citenamefont {Bertholds}\ and\ \citenamefont
  {Dändliker}(1987)}]{bertholds}%
  \BibitemOpen
  \bibfield  {author} {\bibinfo {author} {\bibfnamefont {A.}~\bibnamefont
  {Bertholds}}\ and\ \bibinfo {author} {\bibfnamefont {R.}~\bibnamefont
  {Dändliker}},\ }\href@noop {} {\bibfield  {journal} {\bibinfo  {journal} {J.
  Lightwave Technol.}\ }\textbf {\bibinfo {volume} {LT-5, 7}} (\bibinfo {year}
  {1987})}\BibitemShut {NoStop}%
\bibitem [{\citenamefont {Warken}\ and\ \citenamefont
  {Giessen}(2004)}]{Warken04}%
  \BibitemOpen
  \bibfield  {author} {\bibinfo {author} {\bibfnamefont {F.}~\bibnamefont
  {Warken}}\ and\ \bibinfo {author} {\bibfnamefont {H.}~\bibnamefont
  {Giessen}},\ }\href {\doibase 10.1364/OL.29.001727} {\bibfield  {journal}
  {\bibinfo  {journal} {Opt. Lett.}\ }\textbf {\bibinfo {volume} {29}},\
  \bibinfo {pages} {1727} (\bibinfo {year} {2004})}\BibitemShut {NoStop}%
\bibitem [{\citenamefont {Sumetsky}\ \emph {et~al.}(2006)\citenamefont
  {Sumetsky}, \citenamefont {Dulashko}, \citenamefont {Fini}, \citenamefont
  {Hale},\ and\ \citenamefont {Nicholson}}]{Sumetsky06b}%
  \BibitemOpen
  \bibfield  {author} {\bibinfo {author} {\bibfnamefont {M.}~\bibnamefont
  {Sumetsky}}, \bibinfo {author} {\bibfnamefont {Y.}~\bibnamefont {Dulashko}},
  \bibinfo {author} {\bibfnamefont {J.~M.}\ \bibnamefont {Fini}}, \bibinfo
  {author} {\bibfnamefont {A.}~\bibnamefont {Hale}}, \ and\ \bibinfo {author}
  {\bibfnamefont {J.~W.}\ \bibnamefont {Nicholson}},\ }\href {\doibase
  10.1364/OL.31.002393} {\bibfield  {journal} {\bibinfo  {journal} {Opt.
  Lett.}\ }\textbf {\bibinfo {volume} {31}},\ \bibinfo {pages} {2393} (\bibinfo
  {year} {2006})}\BibitemShut {NoStop}%
\bibitem [{\citenamefont {Wiedemann}\ \emph {et~al.}(2010)\citenamefont
  {Wiedemann}, \citenamefont {Karapetyan}, \citenamefont {Dan}, \citenamefont
  {Pritzkau}, \citenamefont {Alt}, \citenamefont {Irsen},\ and\ \citenamefont
  {Meschede}}]{Wiedemann10}%
  \BibitemOpen
  \bibfield  {author} {\bibinfo {author} {\bibfnamefont {U.}~\bibnamefont
  {Wiedemann}}, \bibinfo {author} {\bibfnamefont {K.}~\bibnamefont
  {Karapetyan}}, \bibinfo {author} {\bibfnamefont {C.}~\bibnamefont {Dan}},
  \bibinfo {author} {\bibfnamefont {D.}~\bibnamefont {Pritzkau}}, \bibinfo
  {author} {\bibfnamefont {W.}~\bibnamefont {Alt}}, \bibinfo {author}
  {\bibfnamefont {S.}~\bibnamefont {Irsen}}, \ and\ \bibinfo {author}
  {\bibfnamefont {D.}~\bibnamefont {Meschede}},\ }\href {\doibase
  10.1364/OE.18.007693} {\bibfield  {journal} {\bibinfo  {journal} {Opt.
  Express}\ }\textbf {\bibinfo {volume} {18}},\ \bibinfo {pages} {7693}
  (\bibinfo {year} {2010})}\BibitemShut {NoStop}%
\bibitem [{\citenamefont {Kurjian}\ \emph {et~al.}(1989)\citenamefont
  {Kurjian}, \citenamefont {Krause},\ and\ \citenamefont {J.}}]{kurkjian}%
  \BibitemOpen
  \bibfield  {author} {\bibinfo {author} {\bibfnamefont {C.~R.}\ \bibnamefont
  {Kurjian}}, \bibinfo {author} {\bibfnamefont {J.~T.}\ \bibnamefont {Krause}},
  \ and\ \bibinfo {author} {\bibfnamefont {M.~M.}\ \bibnamefont {J.}},\
  }\href@noop {} {\bibfield  {journal} {\bibinfo  {journal} {J. Lightwave
  Technol.}\ }\textbf {\bibinfo {volume} {7, 9}} (\bibinfo {year} {1989})},\
  \bibinfo {note} {and references therein}\BibitemShut {NoStop}%
\bibitem [{\citenamefont {Wuttke}\ \emph {et~al.}(2013)\citenamefont {Wuttke},
  \citenamefont {Cole},\ and\ \citenamefont {Rauschenbeutel}}]{Wuttke13b}%
  \BibitemOpen
  \bibfield  {author} {\bibinfo {author} {\bibfnamefont {C.}~\bibnamefont
  {Wuttke}}, \bibinfo {author} {\bibfnamefont {G.~D.}\ \bibnamefont {Cole}}, \
  and\ \bibinfo {author} {\bibfnamefont {A.}~\bibnamefont {Rauschenbeutel}},\
  }\href {\doibase 10.1103/PhysRevA.88.061801} {\bibfield  {journal} {\bibinfo
  {journal} {Phys. Rev. A}\ }\textbf {\bibinfo {volume} {88}},\ \bibinfo
  {pages} {061801(R)} (\bibinfo {year} {2013})}\BibitemShut {NoStop}%
\end{thebibliography}%

\end{document}